\documentclass[aps,pra,twocolumn,secnumarabic,balancelastpage,amsmath,amssymb,
floatfix,showpacs,longbibliography,reprint,superscriptaddress]{revtex4-1}

\usepackage{graphicx} 
\usepackage{color}      
\usepackage{hyperref} 
\usepackage[T1]{fontenc}

\usepackage[UKenglish]{babel}

\begin{document}

\title{Dressed-state electromagnetically induced transparency\\ for light storage in uniform phase spin-waves}

\author{N. \v{S}ibali\' c}
\affiliation{Joint Quantum Center (JQC) Durham-Newcastle, Department of Physics, Durham University, South Road, Durham, DH1 3LE, United Kingdom}
\email{nikolasibalic@physics.org}
\author{J.~M. Kondo}
\altaffiliation[Present address: ]{Departamento de F\'isica, Universidade Federal de Santa Catarina, Campus Trindade, 88040-900, Florian\'opolis, SC, Brazil}
\author{C.~S. Adams}
\affiliation{Joint Quantum Center (JQC) Durham-Newcastle, Department of Physics, Durham University, South Road, Durham, DH1 3LE, United Kingdom}

\author{K.~J. Weatherill}
\affiliation{Joint Quantum Center (JQC) Durham-Newcastle, Department of Physics, Durham University, South Road, Durham, DH1 3LE, United Kingdom}

\date{\today}

\begin{abstract}
We present, experimentally and theoretically,  a scheme for dressed-state electromagnetically induced transparency (EIT) in a three-step cascade system where a four-level system is mapped into an effective three-level system. Theoretical analysis reveals that the scheme provides coherent state control via adiabatic following and provides a generalized protocol for light storage in uniform phase spin-waves that are insensitive to motional dephasing.
The three-step driving enables a number of other features including spatial selectivity of the excitation region within the atomic medium, and kick-free and Doppler-free excitation that produces narrow resonances in thermal vapor. As a proof of concept we present an experimental demonstration of the generalized EIT scheme using the $6S_{1/2} \rightarrow 6P_{3/2} \rightarrow 7S_{1/2} \rightarrow 8P_{1/2}$  excitation path in thermal cesium vapor. This technique could be applied to cold and thermal ensembles to enable longer storage times for Rydberg polaritons.

\end{abstract}

\maketitle

\section{Introduction}

The coherent mapping of light fields into excitations of matter~\cite{Fleischhauer2000a} is a key feature of the storage and manipulation of optical quantum information~\cite{Kimble2008}. Atomic media have significant appeal because they provide inherently narrow and well-defined optical transitions~\cite{Dudin2013}, and are matched to vapour-based single-photon sources. Light fields can be compressed inside a medium by using highly-dispersive transparency windows formed using electromagnetically induced transparency (EIT)~\cite{Lukin2000,Fleischhauer2005}, and then coherently mapped as a atomic excitation through adiabatic following~\cite{Fleischhauer2000a}. Furthermore, the mapping of coherence between an atomic groundstate and a  Rydberg (highly excited) state allows manipulations at the single-photon level~\cite{Lukin2001a} because strong atom-atom interactions lead to very large optical non-linearities~\cite{Pritchard2010,Pritchard2013} producing effective photon-photon interactions that can be used for nonclassical light generation~\cite{Peyronel2012,Firstenberg2013a} and quantum gate protocols~\cite{Paredes-Barato2014,Saffman2016}. However, excitation of Rydberg states typically requires ladder excitation schemes, which produce stored states that are sensitive to atomic motion due to mismatched wavevectors of the excitation lasers.

Atomic motion in atom-based light memories reduces retrieval efficiency, since spatial selectivity of the readout channel crucially depends on constructive interference of contributions from the initially stored spin-wave. This collectively stored excitation inherits its relative phase from the laser beams, with respective wavevectors $\mathbf{k}_i$, used in the storage protocol~\cite{Zhao2008a}. In particular, in ladder excitation schemes, such as those used for Rydberg states~\cite{Mohapatra2007} typically there is a large wavevector mismatch, i.e. $\left| \sum \mathbf{k}_i \right| \equiv 2\pi/\Lambda \neq 0 $, which causes oscillations in the relative phase of the excited medium with a spatial period of $\Lambda$. The thermal motion of the atoms causes the imprinted spin-wave phase grating to diffuse~\cite{Zhao2008a,Firstenberg2013b}, reducing the visibility of the interference pattern that selects the preferential output direction. This effect limits the maximum storage times and minimum dephasing rate for coherent manipulation~\cite{Zhao2008a,Jenkins2012,Dudin2012,Maxwell2013}. 

Here we propose a new approach to writing uniform-phase spin waves, based on a Doppler-free three-photon ladder excitation scheme. With three driving fields oriented in a plane such that the wavevectors cancel, i.e. $\sum \mathbf{k}_i = 0$, each atom picks up the same phase, independent of their spatial position. Therefore the spin-wave written into the highly excited state is insensitive to motional dephasing. In addition, we show that typical two-photon storage protocols can be easily mapped to the four-level schemes, allowing coherent manipulation. The combination of Doppler-free excitation of Rydberg states into collective excitation with uniform phase, and coherent control of atom populations via adiabatic following, highlights some fundamental advantages of three-photon coherent excitation over conventional two-photon schemes. 

The paper is organised as follows. In Section \ref{sec:eitscheme} we introduce the idea of using electromagnetically induced transparency (EIT) occurring due to destructive interference of two excitation paths over a dressed-state excited in a Doppler-free configuration. Subsequently we discuss the important aspects of this scheme for both thermal and cold atom ensembles. In section \ref{sec:unformspinwave} we discuss the scheme as a way of storing light as a collective atomic excitation, forming a spin-wave with uniform phase. We discuss the prolonged storage time and consequences on purifying the readout state down to the single-photon level. In section \ref{sec:coherent_control} we show how one can easily generalise well known 2-photon protocols for light storage through adiabatic following over a dressed state. In section \ref{sec:spatial_selectivity} we discuss the benefits of spatial selectivity afforded by the proposed scheme. Section \ref{sec:experiment} presents a proof-of-principle experimental demonstration of 4-level dressed-state EIT in thermal vapour under Doppler-free conditions. Finally we present conclusions and an outlook in Section \ref{sec:conclusion}.

\section{EIT using an engineered state}\label{sec:eitscheme}

\begin{figure}[b]
\begin{center}
\includegraphics[width=\columnwidth]{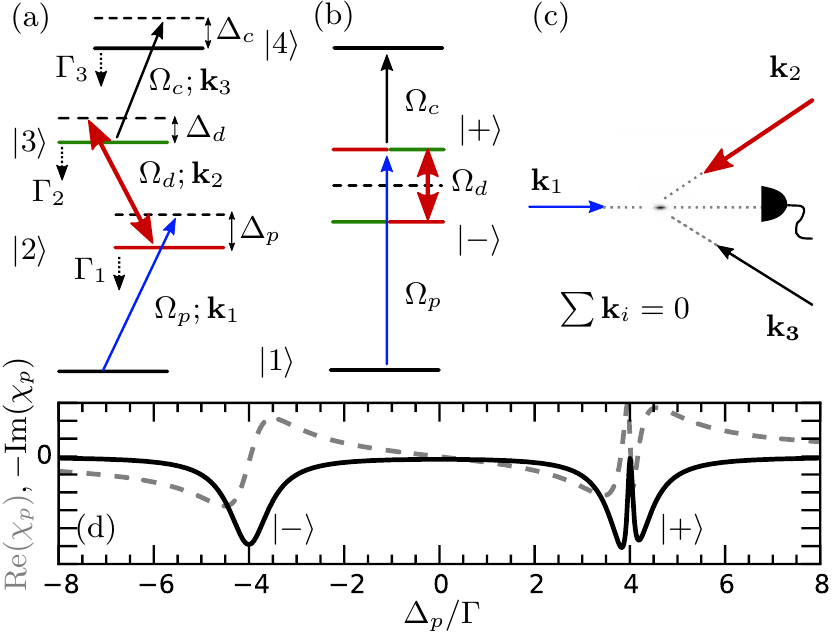}
\end{center}

\caption{\label{fig:levels}(a) Level diagram for bare states in a four-level ladder scheme. (b) Levels in semi-dressed state picture. (c) Three beams are oriented in a plane in the Doppler-free configuration. With both probe and control lasers detuned from bare-state resonance, strong atom-light interactions occur only in the common focus of all the three beams. (d) Theoretical calculation of the real (dashed) and imaginary (solid) parts of the electric susceptibility around the probe transition resonance frequency with parameters $\Gamma_1=\Gamma_2=\Gamma$, $\Gamma_3=0$, $(\Omega_d,\Omega_p,\Omega_c)/\Gamma = (8,~0.1,~0.5)$, $(\Delta_p,\Delta_d,\Delta_c)/\Gamma = (+4,0,-4)$ for a stationary four level atom.}
\end{figure}

We consider a four-level ladder system [Fig. \ref{fig:levels}(a)] driven by three coherent fields (denoted probe, dressing and control) with the corresponding Rabi driving frequencies $\Omega_p$, $\Omega_d$ and $\Omega_c$, and corresponding detunings of the individual lasers of $\Delta_p$, $\Delta_d$ and $\Delta_c$ [Fig.\ref{fig:levels}(a)], as described by the Hamiltonian in $(|1\rangle,\ldots,|4\rangle)$ basis ($\hbar=1$)

\begin{equation}\label{eq:h}
\mathcal{H} = \left( \begin{matrix}
0 & \Omega_p/2 & 0 & 0\\
\Omega_p/2 & -\Delta_p & \Omega_d/2 &0 \\
0 & \Omega_d/2 & -\Delta_p-\Delta_d & \Omega_c/2  \\
0 & 0 & \Omega_c/2 & -\Delta_p - \Delta_d - \Delta_c 
\end{matrix}\right).
\end{equation}

\noindent In addition to the coherent driving, dissipation affecting the system is described by the Lindblad superoperator acting on the density matrix $\hat{\rho}$ of the system $\mathcal{L}[\hat{\rho}] = \sum_k ( L_k \hat{\rho} L^\dagger_k-\frac{1}{2}L_k^\dagger L_k \hat{\rho} -\frac{1}{2}\hat{\rho} L^\dagger_k L_k)$, where we include spontaneous decays with rates $\Gamma_{1\ldots3}$, $L_i = \sqrt{\Gamma_1} |0 \rangle \langle 1 |$, $L_2 = \sqrt{\Gamma_2} |1 \rangle \langle 2|$, $L_3 =  \sqrt{\Gamma_3} |2 \rangle \langle 3|$. The system dynamics are governed by the master equation $\dot{\hat{\rho}} = -i [\mathcal{H},\hat{\rho}] + \mathcal{L} [\hat{\rho}]$.

We focus our attention on the parameter regime where the middle-step laser, resonant with the unperturbed transition $|2\rangle\rightarrow |3\rangle$, $\Delta_d=0$, introduces a strong dressing,  $\Omega_d \gg \Omega_p,\Omega_c$, of the two intermediate states. The goal is to Autler-Townes split the resonance from the ground state as illustrated in Fig.~\ref{fig:levels}(d). The probe and coupling laser fields are both \emph{detuned from the bare-state resonances} ($\Delta_p = -\Delta_c = \Omega_d/2$) so that they fulfil the condition for resonance for one of the dressed states $|+\rangle$ or $ |-\rangle$ [Fig.\ref{fig:levels}(b)]. This engineered state can be used in combination with a strong control $\Omega_c$ and weak probe $\Omega_p$ to open a narrow transparency window for the probe light [Fig.\ref{fig:levels}(d)]. We consider the typical situation where the Rydberg state $|4\rangle$ decay is much weaker compared to that of the intermediate states $\Gamma_3\ll\Gamma_1 \approx \Gamma_2$. To a very good approximation a dark state $|D\rangle$ is formed, which can be obtained from Eq.(\ref{eq:h}) for the limiting worst case of $\Omega_p = \Omega_c$ in the form

\begin{eqnarray*}
|D\rangle &=&\left( |1\rangle - \varepsilon |2\rangle -\varepsilon |3 \rangle + |4\rangle\right)/N,\\
\varepsilon&\equiv&\frac{-\Omega_d+\sqrt{\Omega_c^2+\Omega_d^2}}{\Omega_c},
\end{eqnarray*}

\noindent where $N$ is normalization factor, and $\varepsilon$ characterises the amount of the admixture of the bright (radiatively coupled) states $|2\rangle$ and $|3\rangle$. We see that in the limit of strong dressing, contribution of the bright states $2\varepsilon \sim \Omega_c/\Omega_d\ll 1$ is negligible. This is similar to the double-dark resonance scheme used in $\Lambda$ systems \cite{Lukin1999a}. However, the benefit of using the engineered state for excitation becomes apparent if one considers momentum kick-free, Doppler-free excitation. Whilst typical two-photon driving to highly excited states cannot fulfil the Doppler-free conditions due to mismatch in the excitation laser wavelengths, this is easily done with three driving fields arranged in a plane [Fig.\ref{fig:levels}(c)], further advantages of which will be discussed in the following section.

We note that the dressed-state EIT approach can be extended to more complicated multi-level cascade excitation schemes, following the same principle where probe and coupling laser are detuned from bare-states resonances, so that they are resonant with one of the engineered states that appear in semi-dressed picture of the strong resonant dressing of the intermediate levels. This opens possibilities for using new level-schemes for engineering interactions between the multiple polaritons. However, this will not be considered further in this work.

\section{Uniform phase spin-wave}\label{sec:unformspinwave}

Doppler-free excitation offers additional advantages for atomic memories based on collective superpositions of stored excitations~\cite{Fleischhauer2000a,Duan2001}. Whenever a photon from the probe field is stored as a excitation of the Rydberg state, the atomic excitation picks up a phase from three laser beams. For the Doppler-free configuration of beams with wavevectors $\mathbf{k}_i$, the phase of the excitation is independent of the position $\mathbf{r}$ of the excitation in the cloud since $\sum \mathbf{k_i} \cdot \mathbf{r} = 0$. Since the phase grating determining the output mode will now be determined only by the readout beams, any atomic motion during the storage time does not affect the retrieval efficiency \cite{Jiang2016}. Also, the output mode \emph{direction} can be changed, whilst maintaining the encoded quantum information, by using a different alignment of the output lasers. Similarly, the readout \emph{frequency} can be changed by using a different dressed state for the readout. These features could be important for realizing quantum interconnects \cite{Kimble2008}.

A prominent example where the increase in storage lifetime can have impact is the Rydberg-vapour based single-photon source~\cite{Muller2013}. Recent storage times with light stored as Rydberg excitation in thermal vapour yields lifetime of the memory of only about 1.2~ns~\cite{Ripka2016a}, limited by atomic-motion induced dephasing of the spin-wave imprinted by the two-photon excitation. In order to obtain single photons from the output, one usually relies on strong blockade, assuming that multiple excitations will dephase on the timescale of the excitation laser pulse~\cite{Muller2013,Ripka2016a} due to atom interactions causing level shifts of $C_\alpha/r^\alpha$ for atoms at distance $r$ interacting with resonant dipole-dipole ($\alpha=3$) off-resonant van der Waals ($\alpha=6$) interactions. This requires small excitation volumes in order to achieve strong enough interactions. The longer lifetime, in principle achievable with uniform-phase spin-wave storage, can be used for cleaning the multi-photon output from the output mode after the storage of the initial pulse, since interactions will dephase the spin-waves containing multiple excitations~\cite{Bariani2012}, preventing their readout in the output mode. For clouds of size $l$ this decoupling of the multi-photon excitations from the output mode would happen after $l^\alpha/(C_\alpha h)$, where $\alpha=3$ or $\alpha=6$ depending on whether the Rydberg states are interacting via resonant dipole-dipole interactions or van der Waals interactions. For example, if the excitation is performed in 1~ns, waiting for a time of 100~ns would increase the excitation volume from which we still expect to get only single photon output by factor of 4.6 or 2.1, for dipole-dipole and van der Waals interactions respectively.

\section{Coherent control}\label{sec:coherent_control}

\begin{figure}[t!]
\includegraphics[width=\columnwidth]{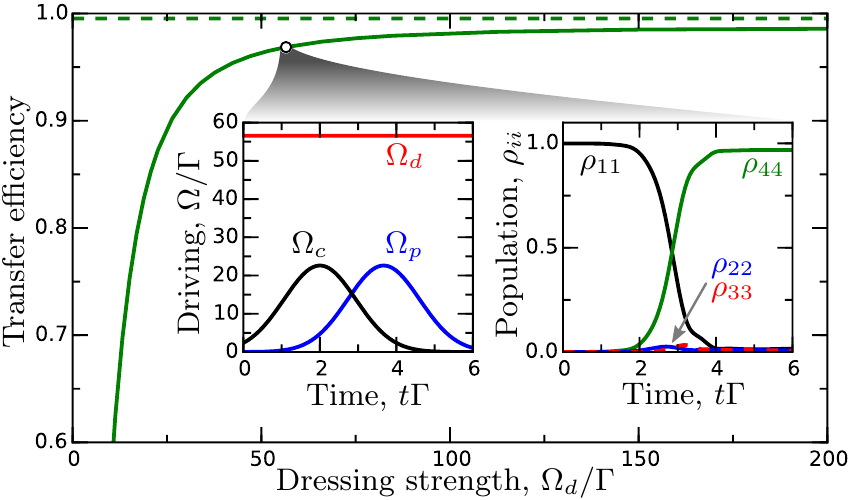}
\caption{\label{fig:stirap}Transfer efficiency to excited state $|4\rangle$ (solid line), compared to that of the three-level scheme (dashed line) for the same control pulses. High efficiency is achieved with strong dressing of the two middle states with generalised STIRAP protocol for four-level ladder scheme (inset). States $|1\rangle$ and $|4\rangle$ are assumed to be long-lived, while the decay constant of each of the two middle states is $\Gamma$.}
\end{figure}

In addition to achieving uniform phase spin-wave excitation, for wider application of the three-photon ladder scheme, coherent manipulation protocols allowing deterministic storage and retrieval \cite{Fleischhauer2000a} are desirable. In the previous section we considered just stochastic excitation. For deterministic, coherent control of populations, off-resonant Doppler-free driving schemes have been proposed \cite{Ryabtsev2011}, however achieving high Rabi driving frequencies is difficult in situations with weak dipole-matrix elements and large detunings. Also, this protocol demands precise control of driving power and timing.

Adiabatic following methods offer a good alternative, relaxing the constraints on precise timing and power, yet allowing deterministic atomic state preparation, as well as mapping of weak quantum fields into atomic excitation of the medium. The standard two-photon STIRAP protocol has been used to prepare atoms in Rydberg states~\cite{Deiglmayr2006}, and can be easily generalized for usage over the engineered state. Details of the protocol are shown in the inset of Fig. (\ref{fig:stirap}). Keeping the dressing laser constantly on, and pulsing the control and the probe laser beams, transfers population between the ground and the Rydberg state [Fig.\ref{fig:stirap}], without populating significantly any of the \emph{two} intermediate states. Two requirements have to be satisfied for efficient transfer: (i) the dressing driving should be stronger than the probe or control driving ($\Omega_d \gg \Omega_c,\Omega_p$)  \cite{Malinovsky1997}; and (ii) the usual three-level STIRAP adiabaticity condition should be satisfied $\Gamma/(T \Omega^2) \ll 1$ \cite{Ivanov2005}, where $\Omega$ is the control ($\Omega_p, \Omega_c$) pulse intensity, $\Gamma$ decay is the constant of the two middle (dressed) states, and $T$ is the characteristic switching time of the two pulses.

The combination of adiabatic following and the existence of a narrow transparency windows, allowing pulse slowing down and compression [Sec. \ref{sec:eitscheme}], suggest that the scheme can be used as a simple generalization of three-level storage protocols \cite{Fleischhauer2000a}, offering above discussed benefits of uniform-phase spin-wave excitation [Sec. \ref{sec:unformspinwave}]. In the limit where multiple Rydberg atoms are excited at distances shorter then their characteristic interaction strength, the protocol can be used for testing proposals that exploit Rydberg-Rydberg interactions in small excitation volumes for state preparation \cite{Petrosyan2013}. Finally, note that the described adiabatic following protocol is only efficient for cold atoms, since in hot atoms the Doppler-effect dephases the system during the adiabatic following, significantly reducing the transfer efficiency. 

\section{Spatial selectivity}\label{sec:spatial_selectivity}

In addition to Doppler-free excitation, the non-collinear orientation of the three beams provides also an excitation volume whose size is determined by the overlap of all three beams. Since both probe and coupling beams are detuned from bare-state resonance, the medium is \emph{transparent} for the beams everywhere except on the overlap region, which is the only place where population of the atoms in the Rydberg or the ground state is changed. The off-resonant, non-collinear, multi-photon excitation scheme [Fig \ref{fig:levels}(c)] allows excitation and probing of well-localized regions in any arbitrary location within the atom medium, whose size can be down to the order of several micrometers if all the beams are tightly focused.

The excitation of atomic vapours confined in spectroscopic cells in this scheme, in combination with narrow Doppler-free features, is promising for electrometry in microwave (MW) and terahertz (THz) domain \cite{Sedlacek2012,Wade2016}, allowing sub-wavelength imaging of fields in the vicinity of structures that are either immersed in the atomic vapour, or are placed next to the spectroscopic cell  \cite{Horsley2015}. The scheme can also provide good spatial resolution for probing of atom-surface interactions \cite{Hinds1997} for patterned surfaces inserted in the vapor cell \cite{Sheng2016}, and exploring non-equilibrium phase transitions \cite{Carr2013b} in small volumes.

In cold atom ensembles, this excitation scheme allows exciting only a fraction of the atomic cloud, which can be important e.g. in Rydberg experiments where one might want to perform excitation within the micrometer range characteristic for Rydberg blockade \cite{Lukin2001a,Petrosyan2013}, within the larger cold atom clouds for state preparation \cite{Petrosyan2013} or for single ion source creation \cite{Ates2013}. Similarly, the scheme can be used for single-site addressing \cite{Weitenberg2011,Labuhn2014,Xia2015,Wang2015,Preiss2015} in 2D and 3D lattices. If the excitation is done for an atomic ensemble held in 2D lattice, the addressing can be done only by moving the dressing laser focus, keeping wide probe and coupling beams, that illuminate the whole lattice, unchanged.

\section{Experimental demonstration of dressed-state EIT}\label{sec:experiment}

\begin{figure}[b]
\includegraphics[width=\columnwidth]{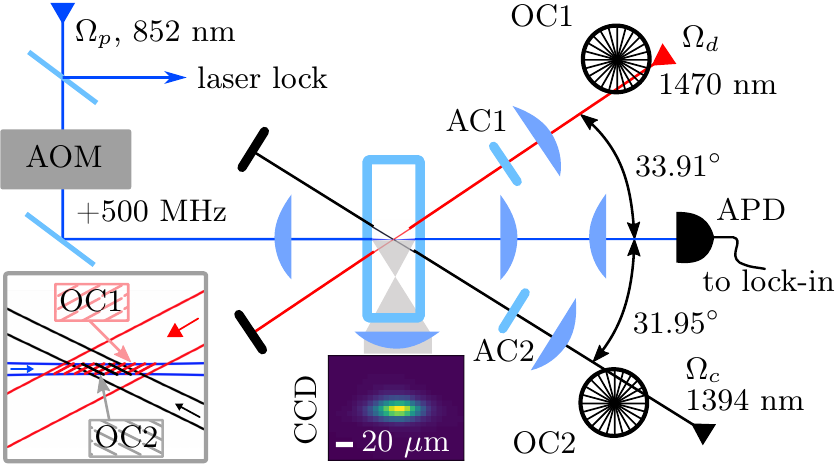}
\caption{\label{fig:exp_setup}Schematic of the experimental setup. An acousto-optic modulator (AOM) provides the frequency offset to the probe beam whose transmission through a 2~mm cesium vapor cell is recorded on avalanche photo-diode (APD). Dressing (1470~nm) and control (1394~nm) beams are passed through achromatic lenses and astigmatism correction plates (AC1 and AC2) before reaching a common focus inside the cell. In the combined focus, the frequency-shifted probe field becomes resonant with the transition which causes strong 852~nm fluorescence that is imaged on the camera (CCD) from the side of the cell. Dynamics within the spot region can be extracted cleanly by performing lock-in detection with one of the optical choppers (OC1 and OC2). Inset on bottom left shows different interaction regions from which dynamics is extracted with corresponding optical choppers.}
\end{figure}

As a proof of concept, we provide a demonstration of the proposed EIT scheme in cesium vapour, using the $6~S_{1/2}~F=4\rightarrow 6~P_{3/2}~F=5 \rightarrow 7~S_{1/2}~F=4 \rightarrow 8~P_{1/2}~F=3,4$ ladder scheme, with corresponding wavelengths 852~nm, 1470~nm and 1394~nm respectively. The first two lasers can be locked on resonance using Doppler-free polarisation spectroscopy~\cite{Pearman2002, Carr2012}. The weak probe beam, set $2\pi\times500$~MHz off-resonance from the transition, is obtained using an acousto-optic modulator in double-pass configuration. The experiment is performed in a 2~mm long quartz vapour cell [Fig.~\ref{fig:exp_setup}] containing cesium vapour at temperature of $50~^\circ$C. All three beams are focused [beam $1/\mathrm{e}^2$ waists $(w_p,w_d,w_c)=(6,28,29)~\mu$m] at a common spot inside the cell. In order to achieve the Doppler-free configuration, dressing and control beams are focused through the 1.25~mm thick quartz windows, at an angle of $32^\circ$ and $34^\circ$ respectively to the direction of the propagation of the probe laser beam which is incident normal to the cell windows. The astigmatism introduced by the windows offsets the focii in sagittal and tangential plane, of both dressing and control beam, by about $\sim0.2$~mm. This is compensated by adding additional quartz windows [AC1 and AC 2 in Fig.~\ref{fig:exp_setup}] of the same thickness and at the same incident angle, but now in sagittal plane, i.e. rotated 90$^\circ$ around propagation direction with respect to the glass window.

\begin{figure}[t!]
\includegraphics[width=\columnwidth]{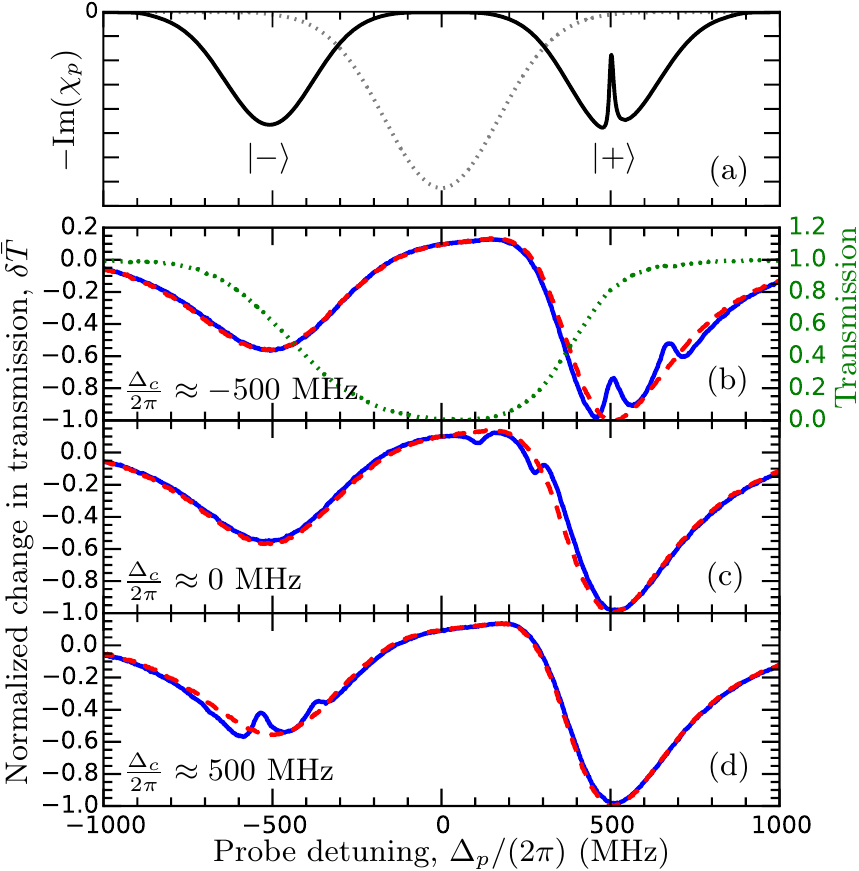}
\caption{\label{fig:exp_th}(a) Four-level theoretical calculation of the imaginary part of the electric susceptibility for the Doppler broadened medium: Dotted (solid) line shows line profile without (with) dressing laser beam $\Omega_d=0$ ($\Omega_d = 2\pi\times 1$ GHz). Solid line shows EIT over one of the dressed states for $\Delta_d = 0$, $\Delta_c=-2\pi\times 500$~MHz for transit lifetime of $\tau=26$~ns. (b-d) Experiment in Cs thermal vapour: Dotted line shows total probe transmission through the cell. Lock-in detection with modulation on the dressing beam shows AT splitting (dashed line). Addition of the control laser (solid line), opens a transparency window when control field is detuned to one of the dressed states (b) and (d). With the control laser on resonance, EIA is observed (c).}
\end{figure}

A theoretical prediction of the probe absorption, excluding the hyperfine splitting, is presented in Fig.\ref{fig:exp_th}(a). Steady state dynamics for the model in Sec. \ref{sec:eitscheme} is calculated for the ensemble of 4-level atoms, taking into account Doppler broadening (at $50~^\circ$C) due to motion of the atoms in the plane defined by the excitation lasers.  Decay rates of the excited states $\Gamma_{1\ldots 3}$ correspond to the natural lifetimes of $6~P_{3/2}$, $7~S_{1/2}$ and $8~P_{1/2}$ respectively. Additionally, each of the states decays to the ground state with the rate $\Gamma_\tau=1/\tau$ due to the finite transit time $\tau$ of the atoms through the excitation region. The transparency peak that opens on one of the dressed states [$|+\rangle$ in Fig.~\ref{fig:exp_th}(a)] is limited now in visibility by the transit time. 

Experimentally obtained level splitting, with the dressing laser beam ($P_d = 4.1$~mW) on $6P_{3/2}~F=5\rightarrow 7S_{1/2}~F=4$ resonance, and the control beam set off-resonance, are presented on Fig.~\ref{fig:exp_th}(b-d). An avalanche photodiode (APD) records the probe beam ($P_d=400$~nW) absorption [Fig. \ref{fig:exp_th}(b) dotted line] through the 2~mm thick vapour region, which includes $\sim 10^2$ times shorter common interaction region. In the interaction region dressing beam induces Autler-Townes (AT) splitting of the $6P_{3/2}\rightarrow 7S_{1/2}$ resonance, which additionally broadens the wings of the Doppler-broadened D2 spectrum. The dressing beam can be switched on and off with an optical chopper [OC1 in Fig.~\ref{fig:exp_setup}]. When the APD signal is demodulated with the lock-in amplifier, we can obtain the change of probe transmission $\delta T$ due to dressing beam in the common interaction region. This reveals two AT peaks [Fig. \ref{fig:exp_th}(b) dashed line], the red-detuned one being smaller due to contribution of other hyperfine states ($F=3,4$) of D2 line. The increased transparency when probe is on resonance is also explained in the dressing picture, with resonance being shifted away from the probe transition due to AT line splitting.  Finally, adding the control laser ($P_c= 8.8$~mW) causes transparency peak to appear when the resonance condition with either of the dressed states ($|+\rangle$ or $|-\rangle$) is achieved [solid line on \ref{fig:exp_th}(b) and \ref{fig:exp_th}(d)]. With maximum absorption in demodulated signal normalized to 1 for maximum absorption, we see that we achieve transparency of $\sim$30\%. Two peaks are observed, corresponding to two hyperfine states $8P_{1/2}~F=3,4$ of the final state, split by $2\pi\times171$~MHz. Note that if the control laser is left on resonance, enhanced absorption is observed [Fig. \ref{fig:exp_th}(c) solid line], which is explained as usual three-photon electromagnetically induced absorption \cite{Ye2002,Carr2012,Kondo2015} .

\begin{figure}[t]
\includegraphics[width=\columnwidth]{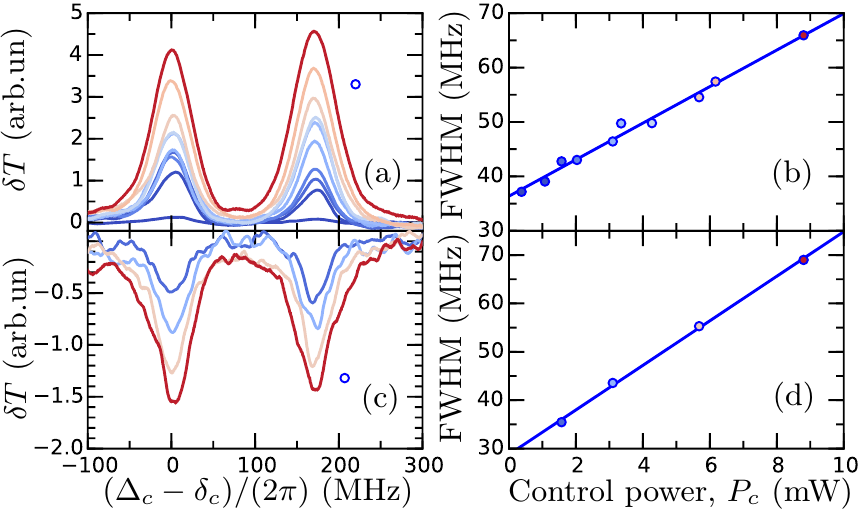}
\caption{\label{fig:linewidths} EIA and EIA under conditions of strong resonant dressing $\Delta_d = 2\pi \times 0$~MHz, $\Omega_d =2\pi\times 1$~GHz. (a) EIT features obtained for off-resonant probe and control $\Delta_p = -\delta_c = 2\pi\times 500$~MHz. (b) FWHM of EIT varies linearly with control power. (c) EIA features are observed when probe and control fields are resonant with bare state transitions $\Delta_p =  -\delta_c \approx 0$. (d) FWHM has linear scaling with control power.}
\end{figure}

To get further insight on the nature of observed EIT resonances, we now scan the control laser, leaving the probe laser locked, with the probe beam tuned $2\pi\times 500$~MHz to the blue of resonance $6S_{1/2}F=4\rightarrow 6P_{3/2}F=5$, and dressing laser locked on resonance $6P_{3/2}F=5 \rightarrow 7S_{1/2}~F=4$. Now the control laser power is modulated with the optical chopper OC2 [Fig.~\ref{fig:exp_setup}], while the dressing beam power is kept constant. Note that with this modulation one probes, in principle, a different spatial part of the interaction region [Fig.~\ref{fig:exp_setup} inset], although in the present case the two regions are almost the same since $w_d\approx w_c$.  The lock-in amplifier demodulated probe absorption signal is presented in Fig.~\ref{fig:linewidths}~(a). Analysis of the resonance (marked with a dot) full-width half-maximum (FWHM) extracted from the Gaussian fits reveals a linear scaling [Fig.~\ref{fig:linewidths}(b)] in accordance with the theory \cite{Erhard2001}. The narrowest features in the limit of $\Omega_c \rightarrow 0$ are about $2\pi\times 36$~MHz. Similar analysis with unlocked lasers, and the probe on resonance $\Delta_p\approx 0 $ yields minimum EIA linewidth of about $2\pi\times 29$~MHz [Fig.~\ref{fig:linewidths}(c-d)]. These linewidths are limited by two factors: (i) the finite time the atoms spent in the interaction region, estimated as a time of flight through the probe beam, that broadens every transition by $\Gamma_t = \bar{v}/\bar{d} \approx 2\pi \times 6$~MHz, where $\bar{v}$ is the average atomic speed, and $\bar{d}=\pi D/4$ average path length through the beam of diameter D (corresponding to probe beam in our case); (ii) averaging of dressing beam power over the region where probe and control beams intersect. The latter can be resolved by using top-hat shaped dressing beam or careful selection of beam overlap. For example, with a dressing beam wider than the probe and coupling beam, one can have a situation where overlap between probe and coupling is probing only the small region of the dressing beam. Dynamics only from that region can be conveniently extracted by modulating the control beam [with OC2 in our setup in Fig.(\ref{fig:exp_setup})]. We note that under similar relative configuration of the beam waists ($w_d>w_p,w_c$), without astigmatism compensation, and with less tight focusing of the probe beam, linewidths of down to $2\pi\times 16$~MHz were observed.

Finally we discuss potential alternative experimental implementations. In rubidium a suitable scheme is $5~S_{1/2} \rightarrow 5~P_{3/2} \rightarrow 4~D_{5/2}\rightarrow n~P,n~F$, and dressing in the middle step can be easily achieved since it  corresponds to the 1529~nm, in the range where erbium-doped fibre amplifiers can provide high power. The non collinear schemes allow Doppler-free excitation for almost arbitrary three- and four- photon excitation schemes, as long as $\sum \mathbf{k}_i = 0$ can be satisfied. In special cases, one can achieve almost complete Doppler-free cancelations working in collinear configurations of the laser beams. For example in cesium $6S_{1/2} \rightarrow 6P_{1/2} \rightarrow 9S_{1/2} \rightarrow n P$ ladder schemes, with corresponding wavelengths of 894~nm, 635~nm, 2.2~$\mu$m allows almost complete Doppler cancellation corresponding to spin-wave of $\Lambda \approx 590~\mu$m. In comparison, the 3-photon scheme in cesium, $6S_{1/2}\rightarrow 6P_{3/2}\rightarrow 7S_{1/2}\rightarrow n P$,  with angle misaligned from perfect Doppler-free condition by 1~mrad, would produce spin-wave with the comparable period of $\Lambda \approx 100~\mu$m. Similarly in lithium $2S_{1/2} \rightarrow 2P_{1/2} \rightarrow 4D \rightarrow nP$ ladder scheme, with corresponding wavelengths of 671~nm, 460~nm, 1465~nm, offers even better Doppler cancelation with corresponding spin-wave period of $\Lambda \approx 1$~mm. Compared to non-collinear schemes, these collinear schemes restrict choice of excitation lasers and associated dipole coupling transition strength. However they are promising for achieving narrowest spectral features of interest in electrometry \cite{Sedlacek2012,Wade2016}, allowing easier addressing of bigger atomic volumes that would reduce transit broadening.

\section{Outlook and conclusion}\label{sec:conclusion}

An intriguing possibility of exploiting the current scheme is in potential new experiments that would have structures placed inside the cells, immersed in the atomic vapour. The spatial excitation and probing selectivity would allow a study of atom dynamics and blackbody decays in confined cavity and waveguide geometries where there is a cut-off in the mode density that modifies the lifetime via the Purcell effect. This is of particular interest for experiments \cite{Goldschmidt2015,Zeiher2016} exploring Rydberg dressing \cite{Henkel2010} and other off-resonant excitation protocols, where blackbody decay is the time-limiting process. There a single blackbody decay to neighbouring opposite parity states, produces impurity that strongly interacts with the rest of the atomic ensemble, bringing the nearby atoms in resonance and triggering resonant excitation avalanche and subsequent loss of atoms from the optical trap. One potential way to extend lifetime of these experiments is by suppressing black-body decay. Cavities with appropriate long-wavelength cut-off can remove the free-space vacuum modes responsible for the decay \cite{Kleppner1981a}, suppressing black-body decay without necessarily using cryogenic cooling. However, the effect of such environment engineering on the strength of atom-atom interactions mediated by near-field virtual photon exchange, with photon energies corresponding to the same free-space modes that cavity suppresses, is not fully understood at the moment. The effectiveness of such solutions potentially could be estimated in the vapour cell experiments, exploring the MW and THz radiation induced modifications of the spectra of the atoms excited in the shielding cavities. These cavities would be immersed in the atomic vapour, with internal shielded excitation region with linear dimensions below the wavelength of the corresponding MW and THz transitions. These experiments would require localisation of atomic excitation on the $\sim10~\mu{\rm m}$ level within the larger vapour cells, which can be provided with the described excitation scheme.

In conclusion, we have proposed and analysed a scheme for multi-level dressed-state electromagnetically induced transparency. Off-resonant excitation of the dressed-state in Doppler-free configuration gives rise to narrow transparency lines, and allows momentum kick-free excitation of Rydberg states. Since both EIT and adiabatic transfer exist, just as in the usual three-level storage protocols, this suggests that this scheme can be used for light storage in four-level cascade schemes, providing the advantage of the collective excitation stored in uniform-phase spin waves. This type of storage should overcome limiting effects of motional dephasing. In addition to the spatial selectivity, this represents one of the fundamental advantages of using more advanced multi-photon excitation schemes \cite{Wade2014,Kondo2015,Lee2015} for coherent control in cascaded excitation configurations.

\begin{acknowledgments}
We thank I.~G.~Hughes for providing comments on the manuscript. This work was supported by Durham University, The federal Brazilian Agency of Research (CNPq), ESPRC (Grant No. EP/M014398/1 and EP/M013103/1).
The data presented in the paper are available \cite{Sibalic2016a}.
\end{acknowledgments}

\bibliographystyle{apsrev4-1}


%

\end{document}